\documentclass[superscriptaddress,prl,twocolumn,nofootinbib]{revtex4-1}

\newcommand{\ket}[1]{|#1\rangle}

\newcommand{\unit}[1]{\ensuremath{\;\mathrm{#1}}}

\newcommand{\be}{\begin{equation}}
\newcommand{\ee}{\end{equation}}
\newcommand{\ba}{\begin{array}}
\newcommand{\ea}{\end{array}}
\newcommand{\bea}{\begin{eqnarray}}
\newcommand{\eea}{\end{eqnarray}}

\usepackage{graphicx}
\usepackage{bm}
\usepackage{amsmath}
\usepackage{color}
\usepackage{float}
\usepackage{hyperref} 

\begin{document}
\title{Quantum-enhanced absorption spectroscopy}
\author{R. Whittaker}
\email[]{Beccie.Whittaker@Bristol.ac.uk}
\author{C. Erven}
\author{A. Neville}
\affiliation{Centre for Quantum Photonics, H. H. Wills Physics Laboratory and Department of Electrical and Electronic Engineering, University of Bristol, Merchant Venturers Building,  Woodland Road, Bristol BS8 1UB, UK.}
\author{M. Berry}
\affiliation{H. H. Wills Physics Laboratory, University of Bristol, Tyndall Avenue, Bristol, BS8 1TL, UK.}
\author{J. L. O'Brien}
\author{H. Cable}
\author{J. C. F. Matthews}
\email[]{Jonathan.Matthews@Bristol.ac.uk}
\affiliation{Centre for Quantum Photonics, H. H. Wills Physics Laboratory and Department of Electrical and Electronic Engineering, University of Bristol, Merchant Venturers Building,  Woodland Road, Bristol BS8 1UB, UK.}

\date{\today}

\begin{abstract}
\noindent
Absorption spectroscopy is routinely used to characterise chemical and biological samples. For the state-of-the-art in absorption spectroscopy, precision is theoretically limited by shot-noise due to the fundamental Poisson-distribution of photon number in laser radiation. In practice, the shot-noise limit can only be achieved when all other sources of noise are eliminated. Here, we use wavelength-correlated and tuneable photon pairs to demonstrate sub-shot-noise absorption spectroscopy. We measure the absorption spectra of spectrally similar biological samples---oxyhaemoglobin and carboxyhaemoglobin---and show that obtaining a given precision in resolution
requires fewer heralded single probe photons compared to using an ideal laser.
\end{abstract}

\maketitle
A sample's absorption spectrum is typically measured by comparing the wavelength and intensity of incident light with the wavelength and intensity of transmitted light. However, the state-of-the-art is bound in precision by the shot-noise limit (SNL) which can ultimately limit precision in practice due to acceptable optical effects on the sample itself from the probe, including damage \cite{wo-natphot-7-28, ta-nphoton-7-229, ta-arxiv:1409.0950}.
Here we demonstrate a statistical benefit to using frequency-correlated photon pairs when performing absorption spectroscopy. By using heralded single photons, there is lower fundamental noise than that of ideal laser emission of equal intensity. This is advantageous in spectroscopy where the error in measurement is optical-power dependent, such as in Doppler thermometry~\cite{stace2010theory,PhysRevA.86.012506}, and for measuring with great precision in short time intervals, while minimising photo-chemistry, such as for observing cell dynamics~\cite{ta-nphoton-7-229}. We measure the spectra of two different types of blood protein---haemoglobin bound to oxygen and haemoglobin bound to carbon monoxide---with sub-SNL precision per detected photon. Resolving two different samples in this way demonstrates capacity to achieve a greater resolution of distinguishing two absorption features in the same sample. We support this by analysing sub-SNL performance when measuring an optical filter.

Correlated pairs of photons can be used to herald the generation of single-photons~\cite{gr-epl-1-173} which can in turn be used to measure optical transmission with precision beyond the SNL~\cite{Jakeman86,br-nphot-4-227}
with optimal performance~\cite{Adesso09}. The increase in precision over an attenuated laser comes from the reduction of possible photon number detection outcomes for correlated photon pairs. Detecting one of the photons in each pair heralds the presence of its counterpart---a single probe photon---which is either absorbed or is not. By contrast, coherent and thermal sources of light contain more than vacuum and single photon contributions, even when attenuated---this leads, at best, to Poisson distributions of the output photon-number detection statistics and error on any subsequent estimate.

The performance of various probe states have been compared using estimation theory for various types of detection schemes~\cite{Hayat99,Adesso09,DAuria06,Monras07}. Fock states $\ket{N}$ have been found to be optimal as probes since they have zero uncertainty of photon-number at the input, resulting in a minimum spread for the photon number at the output~\cite{Jakeman86,Adesso09}.
Increasing $N$ decreases uncertainty in measurement, but does not provide additional advantage in scaling.
This is quantitatively similar to the sub-SNL advantage that is obtainable in entanglement-enhanced phase estimation once losses and decoherence have been accounted for~\cite{es-nphys-7-406}.

Previous experiments that achieve sub-SNL precision to reduce photo-damage to samples~\cite{ta-nphoton-7-229, wo-natphot-7-28} equate to optical phase estimation which relies on non-classical interference via entanglement and squeezing. The quantum-enhancement pursued here is based on the sub-poissonian statistics of single photons~\cite{Jakeman86} that can be achieved equally with single photon emitters or heralded photons generated from parametric processes. This has been achieved for monochromatic absorption imaging using spatially correlated photon pairs~\cite{br-nphot-4-227}. Exploiting frequency correlations, photon pairs have been used to reveal absorption spectra~\cite{sc-apl-83-5560, ka-lpl-5-600, sl-lpl-10-075201,ya-pra-69-013806}---for low illumination, correlated photon pairs enable a high signal-to-noise ratio to be maintained despite increasing levels of background illumination~\cite{ka-lpl-4-722}. However, these spectroscopy experiments did not achieve sub-SNL precision. With increased system efficiency, source brightness and extension of spectral range, the results we present open the way to practical sub-SNL absorption spectroscopy for broad application across science and technology.

Absorption of a sample, $0\leq\alpha\leq1$, is determined by comparing a known input intensity $\bar{N}$ and a measured output intensity $\bar{N}^{\prime}=(1-\alpha)\bar{N}$. This is equivalent to estimating the overall loss $\alpha=1-(1-\alpha_{3})(1-\alpha_{2})(1-\alpha_{1})$, where photon loss occurs (i) before the sample at the source ($\alpha_{1}$); (ii) due to absorption by the sample ($\alpha_{2}$); (iii) after the sample during measurement ($\alpha_{3}$). This treatment is valid, provided $\alpha_{1}$ and $\alpha_{3}$ are calibrated with high accuracy separately. For a dilute solution, $\alpha_{2}$ is related to the absorbance, $A$, of the sample through the Beer-Lambert Law, which gives an exponential attenuation depending on the molar absorption coefficient, $\varepsilon$, the molar concentration, $c$, and the distance through the sample, $l$: $A=\varepsilon cl=ln[1-\alpha_2]$~\cite{BeerLambertLaw}.

Precision in estimating $\alpha$---the reciprocal of the root mean square error $\Delta \alpha$---is limited by the fundamental statistical fluctuations of the input probe due to the quantum nature of light~\cite{gi-sci-306-1330}.  Given $\nu$ repetitions of measurements on the output state $\rho^\prime$ for the output photon number, the precision for estimating $\alpha$ is given by $\Delta \alpha = \Delta_{\rho^\prime}\! \hat{N}/\sqrt{\nu}\bar{N}$, where $\Delta_{\rho^\prime}\! \hat{N}=\sqrt{\langle\hat{N}^2\rangle_{\rho^\prime} - \langle\hat{N}\rangle_{\rho^\prime}^2}$, and $\hat{N}$ is the output photon-number operator.
Absorption spectroscopy currently uses laser emission as its lowest noise input probe, and so the probability to detect $n$ photons from the input probe, $P(n)$, and
$n^\prime$ photons after absorption, $P^\prime(n^\prime)$, are both governed by a Poisson distribution: $P\left( n\right) \mapsto P^\prime \left(n^\prime\right)=e^{-\bar{N}(1-\alpha)} [\bar{N}(1-\alpha)]^{n^\prime}/n^\prime!$. This yields the fundamental limit of precision for classical light (the SNL) $\Delta \alpha_{cl}=\sqrt{(1-\alpha) /\nu \bar{N}}$. Other incoherent broadband probe sources such as sodium lamps, and non-ideal laser setups lead to noisier intensity measurements.  For a Fock state $\vert N \rangle$ acting as the probe, the loss process results in a binomial distribution $P^{\prime }\left( n^\prime\right) =\binom{N}{n^\prime}\left(1-\alpha\right) ^{n^\prime} \alpha^{N-n^\prime}$ which results in $\Delta\alpha_F = \sqrt{\alpha(1-\alpha)/\nu\bar{N}}$, an improvement of $1/\sqrt{\alpha}$ over the SNL. Equivalently, for a given target precision $\Delta \alpha$, a factor of $(1-\alpha)$ fewer single photons in the state $\ket{1}$ are needed than when using a coherent state with $\bar{N}=1$.

\begin{figure}[t!]
  \centering
  \includegraphics[width=0.85\columnwidth]{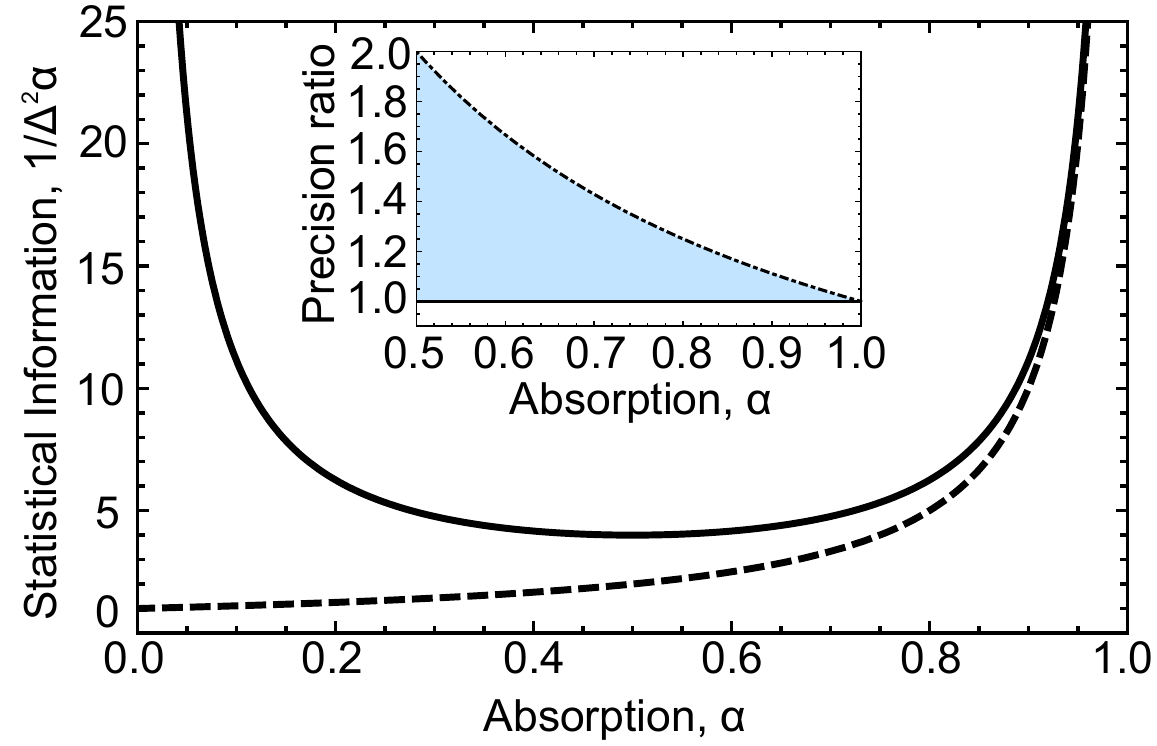}
  \caption{
  \textbf{{Theory
   for precision in estimating $\alpha$}}: use of the Fock state $\ket{1}$ is represented by the solid line and use of a coherent state with the same average intensity of $\bar{N}=1$ photons is represented by the dashed line. Note that divergence at $\alpha=0,1$ corresponds to estimates with vanishing variance---e.g. for $\alpha =1$, repeated trials with $\ket{1}$ will always yield a zero-photon detection event yielding $\Delta^2\alpha = 0$. {Inset}:
   The dot-dashed line corresponds to the ideal quantum advantage that can be obtained using $\ket{1}$, defined by the precision ratio of $\Delta^2\alpha$ for the $\ket{1}$ and a coherent state with $\bar{N}=1$.
   }
\label{TheoryFisher}
\end{figure}
We compare in Fig.~\ref{TheoryFisher} the performance of loss estimation for using the Fock state $\ket{1}$ against using an ideal laser modelled with Poisson-distributed photon-number statistics and having the same photon intensity, $\bar{N}=1$. Performance is quantified using $\nu/\Delta^2\alpha$ which corresponds to the statistical information gained per detected photon---this is the same as the Fisher Information that is widely used in quantum parameter estimation~\cite{Adesso09,Monras07}.
Fig.~\ref{TheoryFisher} illustrates that in principle there is an advantage for using $\ket{1}$ for any $\alpha$, however the magnitude of the improvement scales with $\alpha$ itself: the greatest improvement occurs at low total absorption. Since $\alpha$ is defined including losses throughout the system, this scheme can obtain a quantum advantage with non-perfect components.

\begin{figure}[b!]
  \centering
    \begin{tabular}{l}
    \includegraphics[width= \columnwidth]{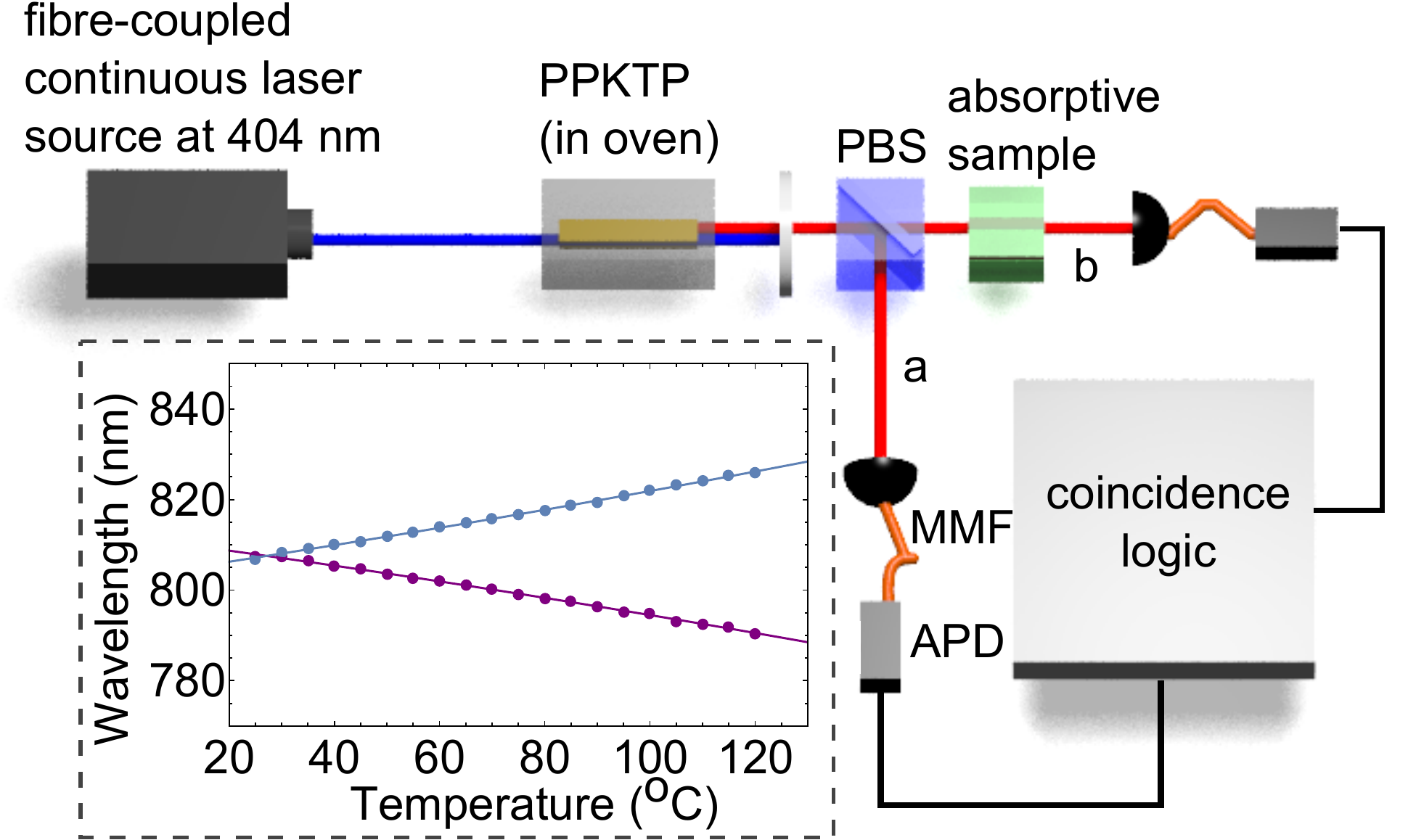}
  \end{tabular}
  \caption{{\textbf{Setup for sub-shot-noise spectroscopy, highlighting the simplicity of the scheme.}}    Wavelength correlated photon pairs are generated using a laser to pump a non-linear optical crystal (PPKTP) that is phase-matched for collinear type-II SPDC and temperature tuned for wavelength control. The sample to be measured can include optical filters, cuvettes containing a liquid chemical or biological sample. PBS-polarising beamsplitter; MMF-multimode fibre; APD-avalanche photodiode.
\textbf{Inset.} {The calibrated temperature dependent joint spectrum of the horizontally polarised (blue) and vertically polarised (purple) photons generated in our PPKTP crystal when pumped with a $403.9$\unit{nm} continuous wave diode laser.
 }
  }
  \label{Fig.Setup1}
\end{figure}

Our experimental setup is shown in Fig.~\ref{Fig.Setup1}(a). A continuous wave diode laser at $403.9$\unit{nm} pumps a 10\unit{mm} long periodically poled potassium titanyl phosphate (PPKTP) non-linear optical crystal, phase-matched for collinear type-II spontaneous parametric down conversion (SPDC), to generate orthogonally-polarised photons in the same spatial mode ($H$ and $V$). These are then separated deterministically using a polarising beamsplitter, while the blue pump laser is removed using three dichroic mirrors normal to the beam, two dichroic mirrors at $45^{\circ}$, one longpass filter and one wide bandpass filter centred at $810 nm$ (FWHM$=50$\unit{nm}).  The photon pairs are generated in a low gain parameter regime of SPDC and so are modelled by $\ket{SPDC}\approx\ket{0}_{a}\ket{0}_{b} + \epsilon \ket{1}_{a}\ket{1}_{b}$.
The crystal is in a temperature-tuneable oven which we use to control the wavelength correlations of the emitted photons---this is pre-calibrated using a single-photon sensitive spectrometer (see inset of Fig.~\ref{Fig.Setup1}). This configuration removes the need to optically determine the wavelength of either photon at the output of our setup during a sample measurement and simplifies the setup to use only two single photon detecting pixels---in our case single photon counting modules and coincidence logic.

Absorption estimates from each single probe photon are distributed across the photon's bandwidth---this is a limiting factor for both the resolution and precision of measured spectra. Here, the bandwidth of photons generated from SPDC is dependent upon that of the pump laser. We use a laser with a linewidth of $0.057$\unit{nm} (FWHM) resulting in linewidths of $\sim 0.5\unit{nm}$ and $\sim 0.7\unit{nm}$ for the photons in arms $a$ and $b$ respectively.  The accessible wavelength range for each arm for our setup is $773 \leq\lambda_{a}\leq 809$ and $806 \leq\lambda_{b}\leq 845$\unit{nm}, restricted by the maximum temperature of our oven $(=200^{\circ}C)$ and our use of a fixed frequency pump laser.
Each path $a$ and $b$ is coupled into multimode fibres to increase photon collection efficiency and to reduce sensitivity to mechanical vibrations.
In the absence of a sample, the efficiency of each arm is $\eta_{a} \sim 35\%$ and $\eta_{b} \sim 29\%$---this includes the $\sim69\%$ specified efficiency of our single-photon detector modules that are used with coincidence logic to record both the singles count rates (\emph{$N_{a}$}, \emph{$N_{b}$}) and the coincidence count rates (\emph{$N_{ab}$}).

\begin{figure}[t!]
  \centering
  \includegraphics[width=0.7\columnwidth]{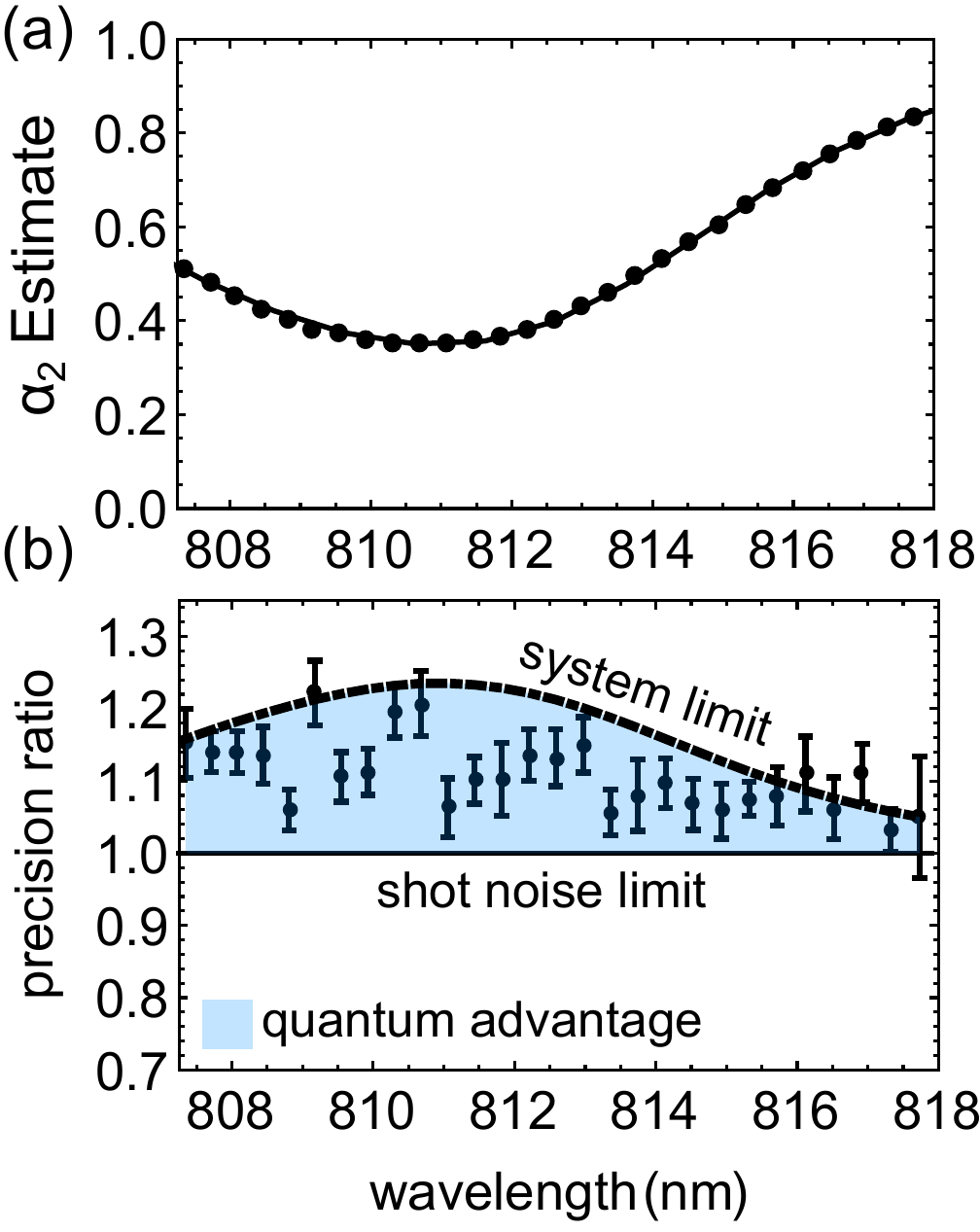}
  \caption{{\textbf{Sub-shot-noise absorption spectra of a control feature.}} A Gaussian bandpass filter was placed in arm $b$ and its spectral response across the wavelength range $807-818$\unit{nm} was measured with correlated photon pairs (dots) and a commercial UV/Vis spectrometer (solid line). (a) The mean absorption of the filter extracted from 1500 estimates for each measured wavelength. (b) Measured quantum advantage (data points) for estimating $\alpha$ of the Gaussian filter, quantified as a ratio of the precision obtainable with an ideal laser (solid line). The theoretical maximum that accounts for our system efficiencies, using ideal quantum states, is computed using $\alpha_{exp}$ and represented by the dot-dashed line. Error bars computed as explained in the main text.}
  \label{Fig.Gaussian}
\end{figure}

Fig.~\ref{Fig.Gaussian} displays the application of our setup to measure the spectral response of a Gaussian bandpass filter~\footnote{(Thorlabs FB$810$-$10$, centre wavelength $810\unit{nm}$, bandwidth $10 \pm 2$\unit{nm} (FWHM))} placed in path $b$. The absorption is estimated directly~\cite{BGN00} from the ratio of heralded single photon detection events in path $a$ to coincidences, according to
\begin{eqnarray}\label{Eq.etaEstExp}
    \alpha _{exp} & = &1- {N_{ab}}/{N_{a}}.
\end{eqnarray}
To take a complete spectrum, $\alpha_{exp}$ is calculated at a range of different crystal temperatures corresponding to a known set of wavelengths. For each probing wavelength we compute the mean value of $\alpha_{exp}=1-(1-\alpha_1)(1-\alpha_2)(1-\alpha_3)$ and the variance $\Delta^2\alpha_{exp}$ over $\nu=1500$ trials using a 1\unit{s} integration time per trial.
To ensure a stable probing wavelength after each temperature change, we enforce a $300$\unit{s} stabilisation period before beginning the next set of measurements. The absorption spectrum of the filter, $\alpha_2$, was found by dividing out the system absorption due to $\alpha_1$ and $\alpha_3$ that was characterised separately.  The mean values of $\alpha_2$ for each wavelength are plotted in Fig.~\ref{Fig.Gaussian}(a) where the black dots are from our experimental setup and the black line is that from a classical scan using a UV/Vis spectrometer. Close agreement verifies comparable accuracy to commercial spectroscopy, up to 0.7nm resolution defined by the photon bandwidth in path $b$. There was a uniform 0.375nm offset between our setup and the spectrometer due to calibration discrepancy between the UV/Vis and the single-photon-sensitive spectrometers, and minor alignment error of the Gaussian filter when mounted in the UV/Vis spectrometer.

The quantum advantage obtainable with our setup to measure the Gaussian filter is quantified in Fig.~\ref{Fig.Gaussian}(b).  We plot the ratio of our computed variance for the heralded single photons, compared to what we would obtain using a shot-noise limited ideal laser (solid line in the inset of Fig.~\ref{TheoryFisher}), for each corresponding mean estimate of $\alpha_{exp}$ in Fig~\ref{Fig.Gaussian}(a). Of the $1500$ trials we use sets of $100$ to compute $15$ quantum advantage parameters; the mean of these are plotted in Fig.~\ref{Fig.Gaussian}(b) and the standard error of each mean value are plotted as the error bars. The blue region shows where a quantum advantage is obtained and the dot-dashed line shows the maximum theoretical advantage for each $\alpha_{exp}$ value. A quantum advantage in precision is achieved across the range of the filter, with a maximum advantage of $22.20\pm0.04\%$ per detected photon.

\begin{figure*}[t!]
  \centering
   \includegraphics[width=1.5\columnwidth]{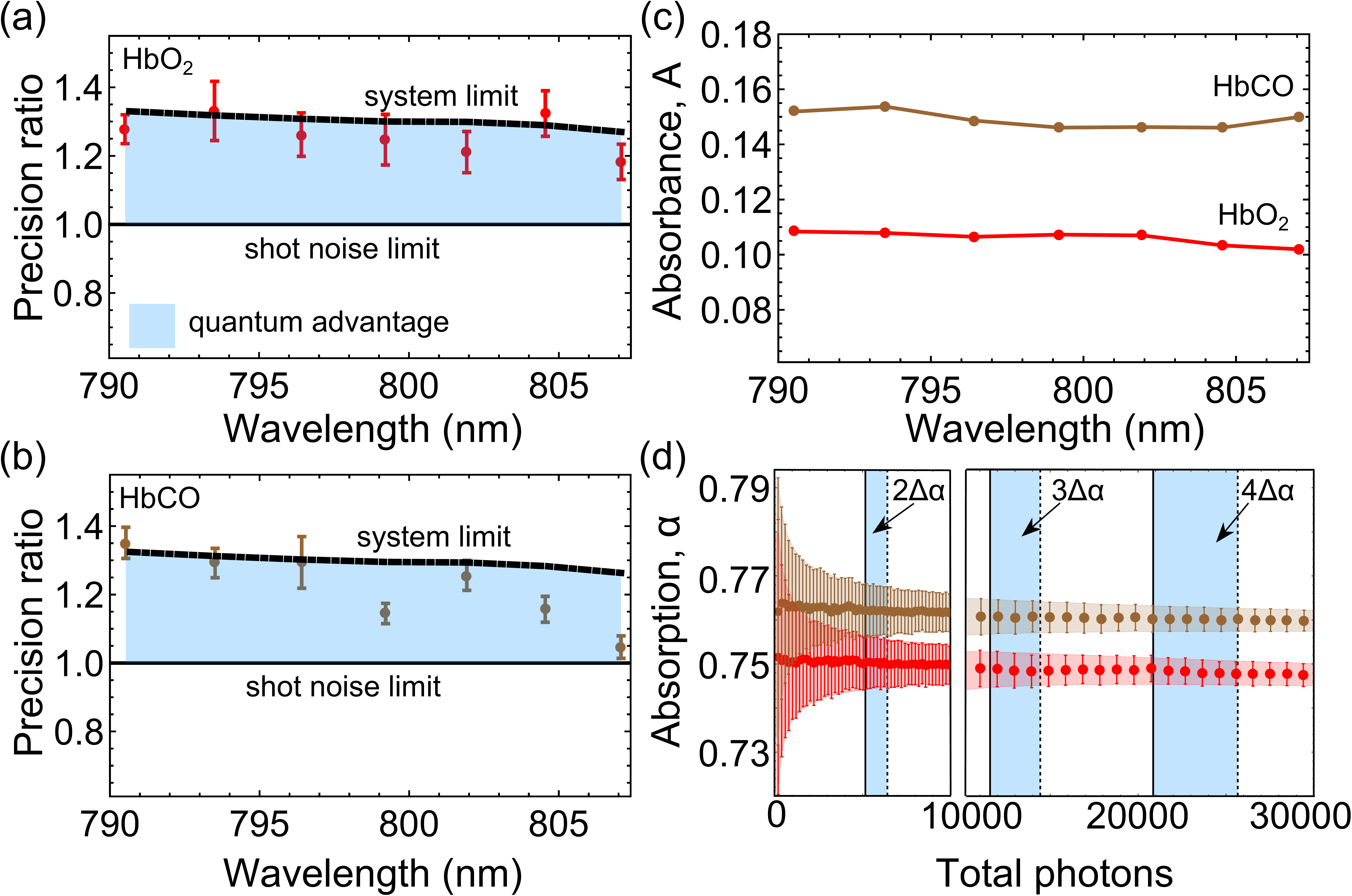}
     \caption{{\textbf{Sub-shot-noise absorbance spectra of HbO$_{2}$ and HbCO.}} Samples of HbO$_{2}$ (red) and HbCO (brown) were placed, in turn, in path $a$ and the spectral response was measured over the wavelength range $\sim 790-808$\unit{nm}. The experimental quantum advantage for estimating absorption $\alpha$ is plotted separately for HbO$_{2}$  (a) and HbCO (b). The error bars are computed using the same treatment as for the Gaussian filter. (c) The mean value of $\alpha$ was converted to absorbance, $A$. (d) Resolving $\alpha_{HbO}$ from $\alpha_{HbCO}$ at $790.5$\unit{nm} probe wavelength.  For each total number of photons, 800 estimates for $\alpha$ are made and the mean is plotted (circles).  The error on these estimates is quantified by one standard deviation ($\Delta\alpha$) of these estimates and plotted as error bars. Solid black vertical lines represent the total number of photons detected in our experiment and required to resolve one $\alpha$ from each other by $2\Delta\alpha$, $3\Delta\alpha$ and $4\Delta\alpha$---these lines are computed using fits to the error bars for each sample. We compare this directly to the total number of photons required to resolve the two $\alpha$ when using an ideal laser (vertical dashed lines). The shaded regions (left to right) correspond to 1250, 2850 and 4810 fewer required photons. We performed a high resolution scan (1ms increments) for the region of $2\Delta\alpha$ separation.
  }
  \label{Fig.HbAbsorbance}
\end{figure*}

We demonstrate sub-SNL spectroscopy of biological samples in the near-infrared region by measuring the absorbance spectra for two different types of haemoglobin: oxyhaemoglobin (HbO$_{2}$) and carboxyhaemoglobin (HbCO). The samples were placed in path $a$ and subjected to the same intensity and data integration time as when measuring the Gaussian filter. From the mean estimate for $\alpha_{2}$, we used the Beer-Lambert law ($A=ln[1-\alpha_{2}]$) to calculate the estimated absorbance spectrum for each sample, plotted in Fig~\ref{Fig.HbAbsorbance}\textcolor{black}{(c)}.  A near-flat spectrum with carboxyhaemoglobin more absorbing than oxyhaemoglobin is obtained, as expected both from literature~\cite{zi-cc-1991,zi-cbp-1997} and from our own characterisation using the UV/Vis spectrometer. We plot in Fig~\ref{Fig.HbAbsorbance}\textcolor{black}{ (a, b)} the quantum advantage in precision of estimating the spectral profile of each sample, showing that a quantum advantage in precision per detected photon is achieved across the entire spectral range for both samples.

Fock states can be used to reduce the number of photons required to discriminate between different absorptions $\alpha$, leading to a higher absorptive resolution than when using an ideal laser.  We demonstrate this in Fig~\ref{Fig.HbAbsorbance}(d) using the two different haemoglobin samples.  Using one probe wavelength ($790.5$\unit{nm}) and a constant intensity, we computed $\alpha$ for an increasing total number of photons---we controlled this by increasing the integration time linearly starting from $1$\unit{ms}, using $1$\unit{ms} increments for high resolution and 5ms increments for lower resolution. For each increment we computed $800$ estimates of $\alpha$ and calculated the mean and standard deviation $\Delta\alpha$.  The standard deviation decreases with increasing total number of photons at a rate that is faster than can be obtained with an ideal laser.  This is shown by the vertical lines in Fig.\ref{Fig.HbAbsorbance} \textcolor{black}{(d)} that quantify the number of photons each scheme requires to resolve $\alpha_{HbO}$ from $\alpha_{HbCO}$ by a separation defined by multiples of the standard deviation, for our experiment (solid) and an ideal laser (dashed).  The difference in total number of photons required in our scheme compared to using an ideal laser will depend on the values of $\alpha$ (c.f. Fig\ref{TheoryFisher}).

We have demonstrated using correlated photon pairs for sub-SNL precision in absorption and absorbance spectroscopy. There are three main aspects of our setup where performance can be improved. \emph{(i)} Increasing overall system efficiency will enable a greater quantum advantage (Fig.~\ref{TheoryFisher})---superconducting detectors are a promising approach, $>90\%$ efficiency have been demonstrated at telecommunication wavelengths~\cite{ma-np-7-210} with promise of high efficiency at shorter wavelengths~\cite{ma-nl-12-4799}; high efficiency single-photon-sensitive cameras~\cite{br-nphot-4-227} could be applied alleviating the need for cryogenic temperatures. \emph{(ii)} Extending emission to a larger spectral range would increase application---this can be accomplished with a tuneable pump laser~\cite{ru-oe-21-10660}. \emph{(iii)} Generating photons with narrow bandwidths would enable higher spectral resolution---for example, cavity-enhancement~\cite{wo-natphot-7-28} would enable application to atomic spectroscopy.

Sub-SNL correlated photon pair spectroscopy is a promising alternative to current forms of spectroscopy---where in practice most systems do not achieve SNL precision---and those where intensity fluctuations have been suppressed~\cite{
ka-josa-17-275,po-prl-68-3020,ka-ol-22-478}. Likely applications include characterising samples that are photo-sensitive or of low concentration or volume---the principle demonstrated in Fig.~\ref{Fig.HbAbsorbance}(d) applied to neighbouring points on a single spectrum, demonstrates the advantage when probing shallow spectral features. Unlike other demonstrations in quantum-enhanced precision measurement, our setup does not require entanglement or multiphoton interference. Equivalent performance could be obtained with extremely narrow bandwidth single photons generated from quantum dots~\cite{yo-jap-101-081711,ja-prl-110-135505} or atom-cavity systems~\cite{ku-prl-89-067901}.  Proposed characterisation of photon-counting and homodyne detectors with Fock states~\cite{ba-ar-1502.00681} could use an iteration of our setup to benchmark spectral response.


\begin{acknowledgments}
\textbf{Acknowledgements:}
The authors would like to thank R. Oulton, A. Young, P. Androvitsaneas, S. Carswell, B. Lang and E. Harbord for use of the spectrometer;  S. Bellamy, P. Dunton and M. Mercer for wet-lab support; G. Magro for materials-lab support; X.Q. Zhou, J. Rarity, M. Padgett, A. Datta, for useful discussions;  N. Tyler, A. Santamato and C. Wilkes for preliminary work on the source.
The authors are grateful for support from DSTL, EPSRC, ERC, NSQI.  JLOB acknowledges a Royal Academy of Engineering Chair in Emerging Technologies. JCFM is supported by a Leverhulme Trust Early Career Fellowship.
\end{acknowledgments}

%
%


%

\twocolumngrid

\end{document}